\documentclass[runningheads,letterpaper]{llncs}

\usepackage[utf8]{inputenc}
\usepackage{amsmath}
\usepackage{amssymb}
\usepackage{stmaryrd}
\usepackage{wasysym}
\usepackage{amsfonts}
\usepackage{graphicx}
\usepackage{graphics}
\usepackage{subfig}
\usepackage{lineno}
\usepackage{url}

\usepackage{clrscode3e}

\DeclareMathAlphabet{\mathpzc}{OT1}{pzc}{m}{it}
      
\newcommand{\idos}{[0, 2\pi)}
\newcommand{\R}[1][2]{\mathbb{R}^{#1}}
\newcommand{\X}[1][\theta]{X_{#1}}
\newcommand{\Y}[1][\theta]{Y_{#1}}

\newcommand{\W}   [2][\theta]{W_{#1}(#2)}
\newcommand{\Wc}  [2][\theta]{\mathcal{W}_{#1}(#2)}

\newcommand{\Vc}  [2][\theta]{\mathcal{V}_{#1}(#2)}

\newcommand{\twedge} [1][\theta]{$#1$-wedge}

\newcommand{\tor}    [1][\theta]{$#1$-orientation}

\newcommand{\rcht}[2][\theta]{\mathcal{RH}_{#1}({#2})}

\newcommand{\ch}[1]{\mathcal{A}(#1)}
\newcommand{\subch}[2][i]{A_{#1}(#2)}
\newcommand{\link}[2]{\ell_{#1,#2}}

\urldef{\mailsa}\path|{alegria_c, garduno_t, areli}@uxmcc2.iimas.unam.mx,|
\urldef{\mailsb}\path|carlos.seara@upc.edu, urrutia@matem.unam.mx|
\newcommand{\keywords}[1]{\par\addvspace\baselineskip
\noindent\keywordname\enspace\ignorespaces#1}

\begin{document}


\title{Rectilinear Convex Hull with minimum area}

\titlerunning{Rectilinear Convex Hull with minimum area}

%
%
\author{%
Carlos Alegr\'{i}a-Galicia\inst{3} \and%
Tzolkin Gardu\~{n}o\inst{3} \and%
Areli Rosas-Navarrete\inst{3} \and\\
Carlos Seara\inst{2} \and%
Jorge Urrutia\inst{1}}
\authorrunning{Rectilinear Convex Hull with minimum area}

\institute{Universidad Nacional Aut\'{o}noma de M\'{e}xico, Instituto
  de Matem\'{a}ticas%
  \thanks{Partially supported by projects MTM2006-03909 (Spain), and
    SEP-CONACYT of M\'{e}xico, Proyecto 80268}%
  \and
  Universitat Polit\`{e}cnica de Catalunya\thanks{Partially supported
    by projects MTM2009-07242, Gen-Cat-DGR2009GR1040}%
  \and
  Universidad Nacional Aut\'{o}noma de M\'{e}xico, Posgrado en Ciencia e\\ Ingenier\'{i}a de la Computaci\'{o}n\\[0.2cm]
\mailsa\\
\mailsb}

%
%

\toctitle{Rectilinear Convex Hull with minimum area}
\tocauthor{Carlos Alegr\'{i}a-Galicia, Tzolkin Gardu\~{n}o,
Areli Rosas-Navarrete, Carlos Seara and Jorge Urrutia}
\maketitle

\begin{abstract}
  Let $P$ be a set of $n$ points in $\R$. We solve the problem of
  computing an orientation of the plane for which the rectilinear
  convex hull of $P$ has minimum area in optimal $\Theta(n\log n)$
  time and $O(n)$ space.
\end{abstract}

\section{Introduction}

The interest in the rectilinear convex hull of planar point sets
arises from the study of
\emph{ortho-convexity}~\cite{ortho-convexity}, also known in the
literature as \emph{rectilinear}, \emph{x-y}, or \emph{orthogonal}
convexity. This non-traditional notion of convexity has been widely
studied since its formalization in the early eighties, and has found
applications in theoretical research fields such as polyhedra
reconstruction~\cite{polyhedra} and fixed point
theory~\cite{fixed-point}, as well as in the study of practical
problems such as digital image shape analysis~\cite{image} and VLSI
circuit layout design~\cite{vlsi}.

A set is said to be \emph{ortho-convex} if its intersection with every
horizontal or vertical line is connected. As it is easy to see, an
ortho-convex set might be non-convex or disconnected. The latter
property hinders the definition of an ortho-convex closure that
resembles the properties of the traditional convex hull and as a
consequence, several definitions have been presented by different
authors~\cite{ottmann}. Throughout this paper, we will use the
\emph{maximal ortho-convex}, or \emph{mr-convex} hull stated by
Ottmann et al.~\cite{ottmann} (see also Matous\v{e}k et
al.~\cite{matousek}), which is defined as follows.

An \emph{orthogonal wedge} is the intersection of two open half-panes
whose supporting lines are orthogonal to each other. Let $\X$ and $\Y$
denote respectively, the $X$ and $Y$ coordinate axes counter-clockwise
rotated by $\theta$ degrees. We call \emph{\tor} to the coordinate
system formed by $\X$ and $\Y$. An \emph{orthogonal \twedge} is an
orthogonal wedge whose supporting lines are parallel to the axes of
the first quadrant of the \tor{} of the plane. Hereafter, let
$P \subset \R$ denote a set of $n$ points in general position (i.e.,
no three collinear points in $P$). An orthogonal wedge is called
\emph{$P$-free} if it does not contain points of $P$.  Let $\W{P}$ be
the set of all $P$-free orthogonal \twedge s, and $\Wc{P}$ be the set
\begin{equation}
 \W{P} \cup \W[\theta + \frac{\pi}{2}]{P} \cup \W[\theta + \pi]{P} \cup
 \W[\theta + \frac{3}{2}\pi]{P}.
\end{equation}

The \emph{Rectilinear Convex Hull} of $P$ is the set
\begin{equation}
\rcht{P} = \R - \underset{w \in \Wc{P}}{\bigcup} w.
\end{equation}

\begin{figure}[ht!]
  \centering
  \subfloat[\label{fig:intro:orientation_1}]{\includegraphics[scale=0.6]{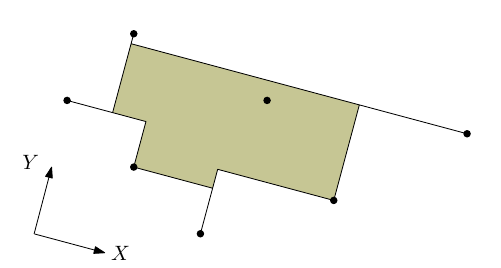}}
  \hspace{1.5cm}
  \subfloat[\label{fig:intro:orientation_2}]{\includegraphics[scale=0.6]{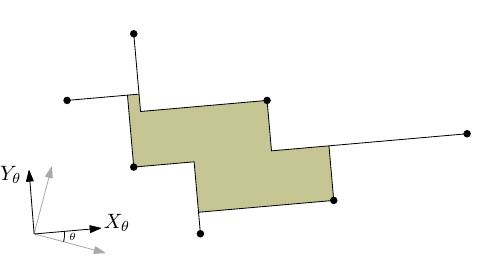}}
  \caption{\label{fig:intro:orientation} $\rcht{P}$ changes along with
    $\theta$.}
\end{figure}

Figure \ref{fig:intro:orientation}\subref{fig:intro:orientation_1}
shows the rectilinear convex hull of a point set. Note in Figure
\ref{fig:intro:orientation}\subref{fig:intro:orientation_2} that the
set $\Wc{P}$ changes along with $\theta$ and thus, $\rcht{P}$ is an
orientation-dependent shape.

From this property, several optimization problems naturally arise for
$\rcht{P}$ when considering the complete set of $\theta$-orientations
of the plane. In this paper we consider the problem of computing a
value of $\theta$ for which $\rcht{P}$ has minimum area. We show that
the set $\{\theta : \theta\in\idos\}$ can be divided into an ordered
set
\[
\{I_1, I_2, \ldots, I_{O(n)}\}
\]
of intervals such that, we can compute the angle $\theta \in I_1$
which minimizes the area of $\rcht{P}$ in $O(n)$ time. We perform the
computation of the angle of optimal area for the subsequent intervals
in constant time. To obtain the set of intervals we use optimal
$\Theta(n \log n)$ time and $O(n)$ space. Our work is based on
techniques from Bae et al.~\cite{coreanos}, Avis et
al.~\cite{theta-maxima}, and D\'{i}az-B\'a\~{n}ez et
al.~\cite{fitting}. Our result improves the $O(n^2)$ time complexity
solution presented recently by Bae et al.~\cite{coreanos}.

\paragraph{Outline of the paper.}

In Section \ref{sec:unoriented-rch} we show how to compute and
maintain $\rcht{P}$ over all the \tor s of the plane. In Section
\ref{sec:minimum-area} we present the algorithm for computing a \tor{}
for which $\rcht{P}$ has minimum area. Finally, in Section
\ref{sec:concluding-remarks} we discuss our concluding remarks.

\section{The Unoriented Rectilinear Convex Hull}
\label{sec:unoriented-rch}

In this section we present an algorithm to compute and maintain
$\rcht{P}$ over all $\theta$-orientations of the plane in
$\Theta(n \log n)$ time.

\subsection{Preliminaries}
\label{subsec:preliminaries}

A \emph{vertex} of $\rcht{P}$ is a point in $P$ that lies on the
boundary of $\rcht{P}$.  For any fixed $\theta$, the computation of
$\rcht{P}$ has a close relation to the \emph{Set Maxima
  Problem}~\cite{preparata-maxima,preparata-comp_geom}, since the set
of vertices of $\rcht{P}$ is equal to the set $\Vc{P}$ of maximal
points of $P$ with respect to vector dominance in the four quadrants
defined by $X_\theta$ and $Y_\theta$~\cite{coreanos,ottmann}.

Since we can specify $\rcht{P}$ using $\Vc{P}$, it is possible to keep
track of the changes in the set of vertices of $\rcht{P}$ over all
\tor s in the plane, by maintaining $\Vc{P}$ as the plane rotates in a
counter-clockwise direction. The maintenance of this set can be done
using the results given in~\cite{theta-maxima, fitting} in the manner
we describe bellow.

\begin{figure}[ht!]
  \centering
  \includegraphics[scale=0.8]{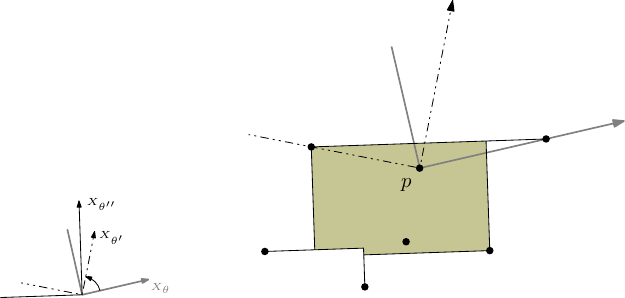}
  \caption{A set $P$ and
    $\rcht[\theta'']{P}$.\label{fig:prel:interval}}
\end{figure}

We call \emph{apex} of an orthogonal wedge to the intersection point
of its supporting lines.~Every point $p \in \Vc{P}$ is the apex of a
$P$-free orthogonal $\theta$-wedge contained in $\Wc{P}$. Figure
\ref{fig:prel:interval} shows us a point $p$ that is vertex of
$\rcht{P}$ in the interval $I_p=[\theta,\theta')$, and the orthogonal
$\theta$ and $\theta'$-wedges with apex in $p$.  The endpoints of
$I_p$ mark the \emph{in}- and an \emph{out}-events of $p$, i.e., the
values of $\theta$ when $p$ belongs and stops being in
$\Vc{P}$. Moreover, a point can be the apex of an orthogonal
$\theta$-wedge during at most three intervals of \tor s.

\begin{theorem}[D\'{i}az-Ba\~{n}ez et al.~\cite{fitting}]
  The set of intervals of orientations in which the elements of $P$
  are maximal with respect to vector dominance, and the ordered set of
  in- and out-events, can be computed in optimal $\Theta(n\log n)$
  time and $O(n)$ space.
\end{theorem}

\begin{figure}[ht!]
  \centering
  \subfloat[\label{fig:prel:start_event}Full interval start]{\includegraphics[scale=0.9]{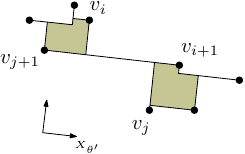}}
  \enskip{}
  \enskip{}
  \subfloat[\label{fig:prel:middle}]{\includegraphics[scale=0.9]{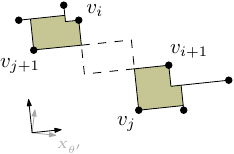}}
  \enskip{}
  \enskip{}
  \subfloat[\label{fig:prel:stop_event}Full interval end]{\includegraphics[scale=0.9]{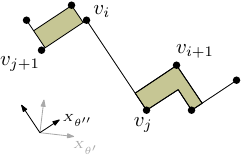}}
  \caption{An overlap disconnects $\rcht{P}$. The ends of the overlap
    full interval.\label{fig:prel:overlap}}
\end{figure}

Let us denote by $V_\theta(P)$ to the set of maximal points of $P$ in
the first quadrant defined by $X_\theta$ and $Y_\theta$.  Since
$\rcht{P}$ is an ortho-convex set, it is monotone with respect to
$X_{\theta}$. Thus, the points of $P$ in $V_\theta(P)$ can be
relabelled as $v_1,\cdots,v_m$ in increasing order according to
$X_{\theta}$. Let $w_i(\theta)$ denote the $P$-free orthogonal
$\theta$-wedge supported by two consecutive points
$v_i,v_{i+1} \in V_{\theta}(P)$. We say that two wedges $w_i(\theta)$
and $w_j(\theta + \pi)$ are \emph{opposite} to each other. Note in
Figure \ref{fig:prel:middle} that, if the intersection of
$w_i(\theta)$ and $w_j(\theta + \pi)$ is non empty, then $\rcht{P}$ is
disconnected. In such case we say that $w_i(\theta)$ and
$w_j(\theta + \pi)$ \emph{overlap}, and denote its intersection as
$O_{\theta}(i,j)$.

Let $\mathcal{O}_\theta(P)$ be the set of all overlaps
$O_{\theta}(i,j)$ of $\rcht{P}$ in the $\theta$-orientation of the
plane. It is easy to see that $\mathcal{O}_\theta(P)$ can be computed
from $V_{\theta}(P)$ in $ O(n)$ time. Thus, $\rcht{P}$ can be computed
in $\Theta(n\log n)$ time and $O(n)$
space~\cite{preparata-maxima,preparata-comp_geom} for a fixed value of
$\theta$.

To maintain the set $\mathcal{O}_{\theta}(P)$ we will compute an
ordered list of angles that resembles the one computed for the in- and
out- events. Since overlap events do not necessarily match up with in-
or out-events~\cite{coreanos}, they have to be computed independently.

\subsection{Overlaps characterization}

\begin{figure}[ht!]
  \centering
  \subfloat[\label{fig:prel:arcchain_in}In-event]{\includegraphics[scale=1]{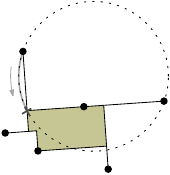}}
  \enskip{}
  \enskip{}
  \enskip{}
  \enskip{}
  \subfloat[\label{fig:prel:arcchain_interior}]{\includegraphics[scale=1]{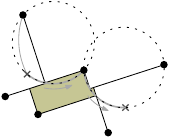}}
  \enskip{}
  \enskip{}
  \enskip{}
  \enskip{}
  \subfloat[\label{fig:prel:arcchain_out}Out-event]{\includegraphics[scale=1]{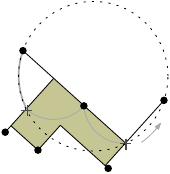}}
  \caption{\label{fig:prel:arcconstruct} The apexes path as the \tor \
    changes. }
\end{figure}

As the $\theta$-orientation of the plane \emph{rotates}
counter-clockwise from $0$ to $2\pi$, the apex of every $P$-free
orthogonal $\theta$-wedge supported by at least two points in $\Vc{P}$
traces a circular arc that can be oriented as shown in
Figure~\ref{fig:prel:arcconstruct}.

\begin{definition}
  The \emph{arc-chain} $\mathcal{A}(P)$ of $P$ is the closed curve
  formed by the union of the set of arcs traced by the apexes of the
  wedges supported by at least two elements of $P$ which belong to
  $\Vc{P}$ for some $\theta \in [0, 2\pi)$.
\end{definition}

\begin{figure}[ht!]
  \centering
  \includegraphics[scale=0.7]{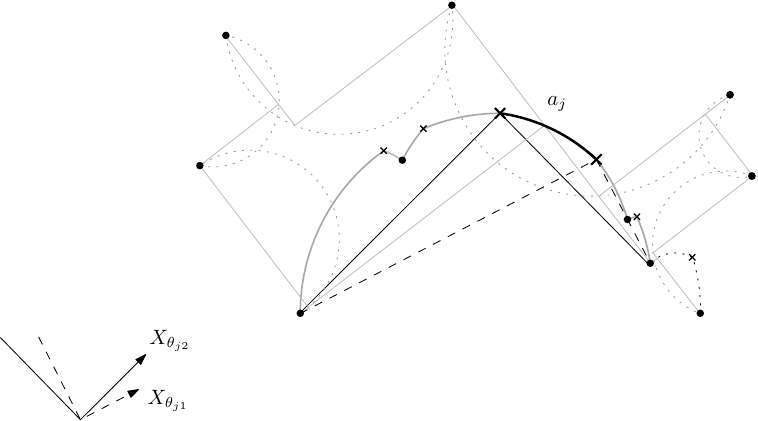}
  \caption{The interval of orientations of an
    arc.\label{fig:prel:arc_interval}}
\end{figure}

The arc-chain can be computed using the in- and out-events list in
$O(n)$ time, since every arc is a segment of a circumference with
diameter in consecutive points of the sets
$V_\theta, V_{\theta+\frac{\pi}{2}},V_{\theta+\pi}$ and
$V_{\theta+\frac{3\pi}{4}}$, and for every arc, its ends correspond to
the in- and out-events of one of the points determining the diameter
of the circumference to which the arc belongs (see Figures
\ref{fig:prel:arcchain_in} and \ref{fig:prel:arcchain_out}). Thus,
each arc can be associated with an interval of orientations in which,
the wedge that traces the arc, is part of the rectilinear convex hull
(see Figure \ref{fig:prel:arc_interval}).

\begin{figure}[ht]
  \centering
  \subfloat[\label{fig:prel:arc_chain_1}]{\includegraphics[width=0.45\textwidth]{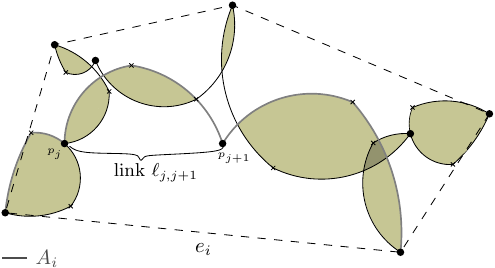}}
  \enskip{}
  \enskip{}
  \enskip{}
  \subfloat[\label{fig:prel:arc_chain_2}]{\includegraphics[width=0.45\textwidth]{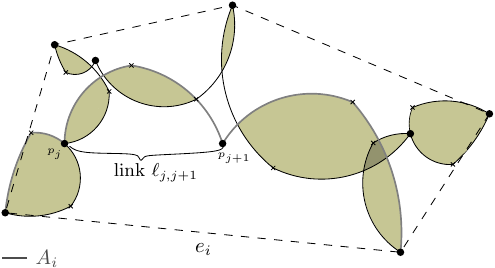}}
  \caption{The \emph{arc-chain} $\ch{P}$ of $P$.}
  \label{fig:prel:arc_chain}
\end{figure}

Let $\{e_1,\cdots,e_h\}$, be the set of edges of the convex hull of
$ P $ in counter-clockwise order. A \emph{subchain}
$\subch{P} \in \ch{P}$ is a subsequence of consecutive elements of
$\ch{P}$, whose endpoints are the endpoints of a convex hull edge
$e_i$.  It is easy to see that $\subch{P}$ is monotone in the
direction determined by $e_i$. Thus, the orthogonal projection of
$\subch{P}$ on $e_i$ defines a total order ($\prec_i$) on the set of
endpoints of its arcs.  Moreover, using the convex hull edges and
($\prec_i$), it is possible to label the arcs in $\ch{P}$ starting by
the leftmost point of $P$ in counter-clockwise order, in such a way
that $\ch{P} = \left\langle a_1,\cdots,a_l\right\rangle $
(Figure~\ref{fig:prel:arc_chain_1}). Let us consider that the ends of
the arcs in any subchain $\subch{P}$ are labeled as $p_1, \ldots, p_t$
such that if $r<s$, then $p_r \prec_i p_s$. A \emph{link}
$\link{r}{s}$ is a subsequence $p_r, \ldots , p_s$ of $\subch{P}$ such
that for $r<t<s$, $p_r,p_s \in P$, $p_r \prec_i p_t \prec_i p_s$ and
$p_t \notin P$.

Now we characterize an overlap using the arc-chain.  Consider four
points $v_i, v_{i+1} \in V_\theta(P)$ and
$v_j,v_{j+1} \in V_{\theta+\pi}(P)$ that support two overlapping
wedges (Figure \ref{fig:prel:overlap}). Let us observe that if two
opposite wedges in $\Wc{P}$ overlap, since these wedges are $P$-free,
the arcs traced by their apices belong to \emph{links} that intersect
(see Figure~\ref{fig:prel:arc_chain_1}). Now, the overlap between the
wedges $w_i(\theta)$ and $w_j(\theta+\pi)$ is possible only in an
interval $[\theta', \theta'']$ where the $\theta'$-orientation is
parallel to the segment $\overline{v_{j+1}v_{i+1}}$ and in
$\theta''$-orientation the $Y$-axis is parallel to the segment
$\overline{v_iv_j}$, see Figures \ref{fig:prel:start_event} and
\ref{fig:prel:stop_event}, respectively. We call \emph{full overlap
  interval} to this interval of orientations and we say that this
interval is a proper interval when $\theta' \leq \theta''$.

Any arc $a_j$ corresponds, for some interval
$I_j= [ \theta_{j1}, \theta_{j2} ]$, to a particular $P$-free
orthogonal $\theta$-wedge, $w_i ( \theta )$ for every $\theta \in I_j$
(Figure \ref{fig:prel:arc_interval}). It is important to notice that
$I_j$ is different from the intervals of the points that support
$w_i ( \theta )$, indeed $I_j$ is contained in the intersection of the
intervals of these points.

Since an overlap is possible only between opposite wedges, in order to
determine if two given arcs $a_j, a_k$ admit opposite wedges, we
define the \emph{facing interval} for them as the interval
$I_f= [\theta_{j1} + \pi, \theta_{j2 + \pi}] \, \cap \, [\theta_{k1},
\theta_{k2}]$.

\begin{lemma}{\label{lem:overlap:conditions}(An overlap interms of arcs)}
  Given two arcs $a_j$ and $a_k$, with its corresponding intervals
  $I_j=[\theta_{j1},\theta_{j2}]$ and
  $I_k=[\theta_{k1},\theta_{k2}]$. Let $w_q(\theta)$ denote the
  orthogonal $\theta$-wedge whose apex traces $a_j$ for
  $\theta \in I_j$ and let $w_r(\theta)$ denote the respective wedge
  of $a_k$ for any $\theta \in I_k$. This wedges overlap if and only
  if the full overlap interval is proper, the facing interval is not
  empty and the intersection between the full overlap interval and the
  facing interval is a proper interval.
\end{lemma}

Then, if there is an overlap between the wedges of two given arcs, the
interval of the overlap is the intersection between the full interval,
the facing interval, and the intervals of both arcs.

Let $I_{i,j}$ be the the interval of $\theta$ for which $w_i(\theta)$
and $w_j(\theta+ \pi)$ overlap. We call the ends of $I_{i,j}$ the
\emph{start}- and \emph{stop}- events of $O_{\theta}(i,j)$. The thick
arc segments show the overlap interval in
Figure~\ref{fig:prel:arc_interval}).

\begin{figure}[ht!]
  \centering
  \includegraphics[scale=0.5]{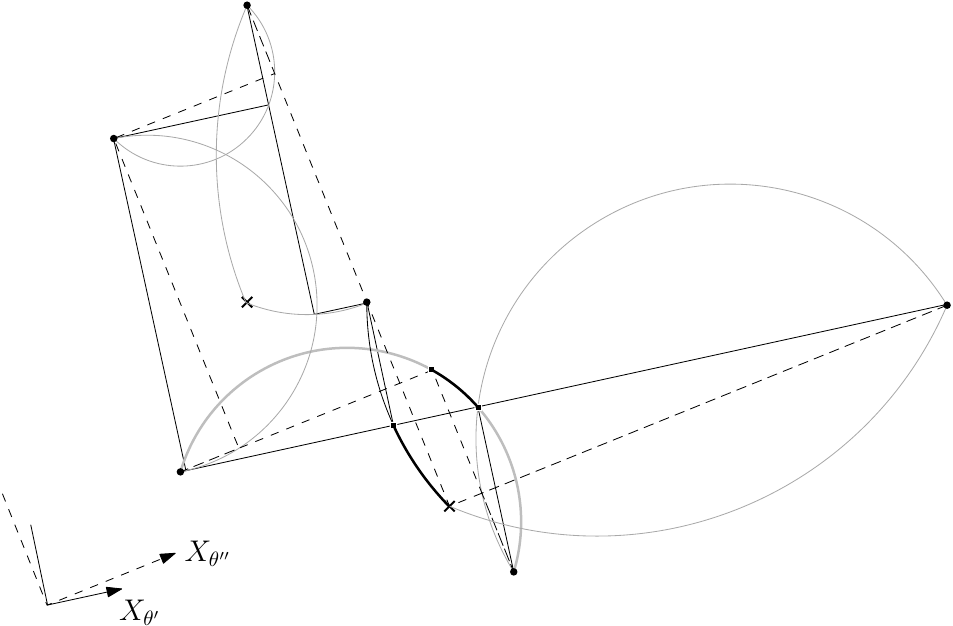}
  \caption{The interval of orientations of an overlap
    $I_{i,j}=(\theta',\theta'')$.\label{fig:prel:overlap_interval}}
\end{figure}

Once we have characterized the overlaps, let us notice that if an
overlap takes place, then the links that contain the arcs involved in
the overlap had intersected. In consequence, it is possible to compute
the overlaps and its corresponding events by computing the
intersection points between links, but before using the intersections,
it is critical to show that there are $O(n)$ intersection points
between links in order to keep the algorithm subquadratic. The
following paragraphs show some properties that help us to show the
linear number of intersections between links in the arc chain.

Using the fact that every point in any subchain $\subch{P}$ is an apex
of an orthogonal $P$-free $\theta$-wedge, the following lemma follows
easily:

\begin{lemma}\label{lem:angle}
  Let $a,b,c \in \subch{P}$ such that $a\prec_i b\prec_i c$. Then,
  $\frac{\pi}{2}\leq \angle abc <\pi$.
\end{lemma}

As a consequence of Lemma \ref{lem:angle}, any link $\link{r}{s}$ is
contained in the disk that has the segment $\overline{p_rp_s}$ as
diameter (the link-disk $D_{r,s}$), thus the number of links that
intersect the link-disk is greater than or equal to the number of
links that intersect $\link{r}{s}$. We denote by $d(r,s)$ to the
diameter of $\link{r}{s}$.

Let $\mathcal{D}(P)=\{D_0,D_1,\ldots,D_h\}$ be the set of disks such
that $D_i$ has the convex hull edge $e_i$ as a diameter and radius
$r_i$. Using a result given in \cite{theta-maxima} we show the
following lemma.

\begin{lemma}\label{lem:chdisks}
  Given $p,q \in P$, the disk that has $\overline{pq}$ as diameter and
  radius $r$ is intersected by at most 28 disks
  $D_i \in \mathcal{D}(P)$ such that $r_i \geq r$.
\end{lemma}

Into each of the disks in $\mathcal{D}(P)$ that intersect a link-disk
$D_{r,s}$, there is a constant number of links with diameter greater
than $d(r,s)$ that intersect $\link{r}{s}$.

\begin{figure}[ht!]
  \centering
  \includegraphics[scale=0.5]{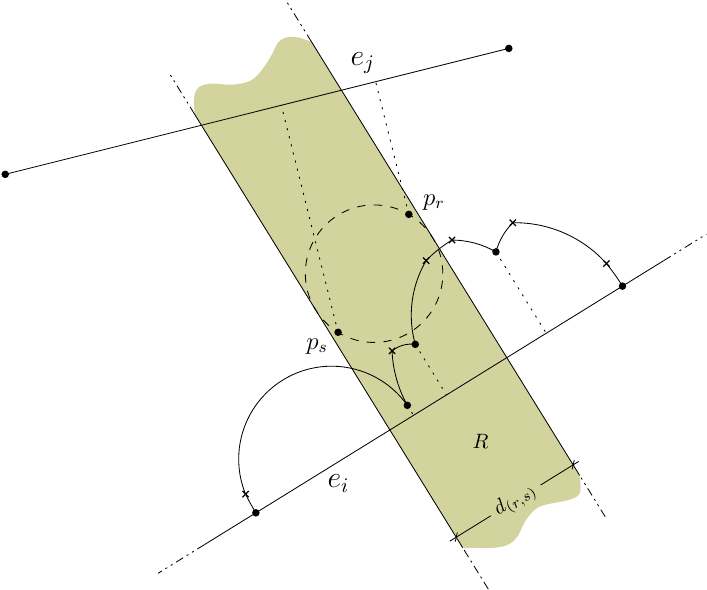}
  \caption{Construction to count the links in $D_i$ that intersect
    another link $\link{r}{s}$ \label{fig:prel:intersect_count1}}
\end{figure}

\begin{lemma} \label{lem:linkInter}
  Let $D_i \in \mathcal{D}(P)$ be a disk of radius $r_i$ that
  intersects some given link disk $D_{r,s}$ of radius $r < r_i$. There
  are at most five links in $D_i$ that intersect $\link{r}{s}$ with
  diameter greater than $d(r,s)$.
\end{lemma}

\begin{figure}[ht]
  \centering
  \subfloat[\label{fig:prel:subch_int2}]{\includegraphics[width=0.45\textwidth]{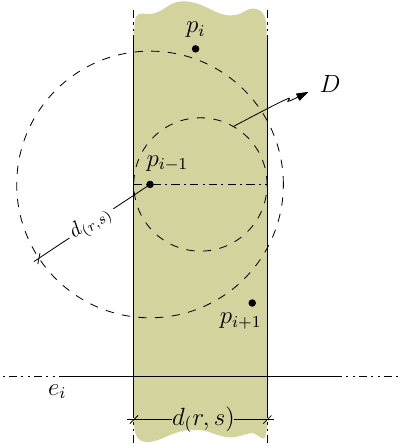}}
  \enskip{}
  \enskip{}
  \enskip{}
  \subfloat[\label{fig:prel:subch_int3}]{\includegraphics[width=0.45\textwidth]{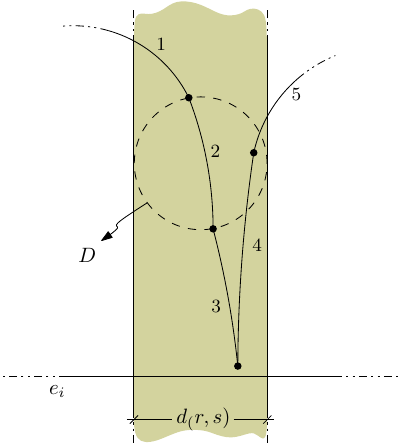}}
  \caption{A link $\link{j}{k}$ can be intersected by at most five
    links contained in some $D_i \in \mathcal{D}(P)$.}
  \label{fig:prel:links_disposition}
\end{figure}

\begin{proof}
  Let $R$ be the closed region bounded by the lines that project
  $D_{r,s}$ on the line that crosses the endpoints of $e_i$ (see
  Figure \ref{fig:prel:intersect_count1}), since subchains are
  monotone, then the links in $\subch{P}$ that intersect $D_(r,s)$
  must have at least one endpoint in $R$.

  Now let us consider the links of $\subch{P}$ with endpoints in
  $R$. For simplicity we suppose that all these links have greater
  diameter than $d(r,s)$. It is easy to see that in order to have many
  links into region $R$, the endpoints of the links should be placed
  in the configurations showed in Figure
  \ref{fig:prel:links_disposition}. The configuration showed by Figure
  \ref{fig:prel:subch_int2}, considers three points
  $p_{i-1},p_{i},p_{i+1}$ such that $p_i$ is a \emph{peak} (i. e. the
  distances from the other points to $e_i$ are smaller then the
  distance between $p_i$ and $e_i$). In this case, with the help of a
  circle with diameter parallel to $e_i$ and lenght $d(r,s)$ that
  passes through $p_{i-1}$ it is possible to show that if $p_i$ must
  be at distance grater than $d(r,s)$ from $p_{i-1}$ and from
  $p_{i+1}$ then $p_i$ cannot be part of a link in $\subch{P}$, since
  the angle $\angle p_{i-1}p_ip_{i+1}$ would be less than
  $\frac{\pi}{2}$, contradicting Lemma \ref{lem:angle}.

  Since there are no crests into region $R$, there is at most one
  \emph{valley}, like the one showed in Figure
  \ref{fig:prel:subch_int3}, therefore there are at most five links in
  $\subch{P}$ that intersect $\link{r}{s}$.
\end{proof}

The open area bounded by $\subch{P}$ and $e_i$ is $P$-free since it is
covered by $P$-free $\theta$-wedges. Thus, two intersecting links have
at least two intersection points, and by Lemma \ref{lem:angle} this
number is tight, as none of the intersecting links can cross a line
segment joining its intersection points.

As a consequence of lemmas \ref{lem:angle}, \ref{lem:chdisks} and
\ref{lem:linkInter}, we have:

\begin{lemma}
  Any link $\link{j}{k}$ is intersected at most 280 times by links of
  diameter greater than or equal to $d(j,k)$.
\end{lemma}

Considering the previous results, and that we have a linear number of
links, we have the following result that is a central tool for
computing the start- and stop-events list in $O(n\log n)$ time.

\begin{theorem}\label{thmLin}
  There are $O(n)$ intersections between links in $\mathcal{A}(P)$.
\end{theorem}

\subsection{Computing the start- and stop- events}

Since in the previous section it has been proved that there are a
linear number of intersection points between arcs, in this section we
describe how to compute the start- and stop- overlap event list.

Let us color the arcs in $\ch{P}$ in the following way: red if the arc
belongs to a subchain that corresponds to an edge in the upper chain
of the convex hull, and the arc will be blue otherwise. Then we
compute the bichromatic intersections using the Bentley and Ottmann
plane sweep algorithm \cite{Bentley-Ottman} in $O(n \log n)$.

\begin{figure}[h!]
  \centering
  \includegraphics[width=0.5\linewidth]{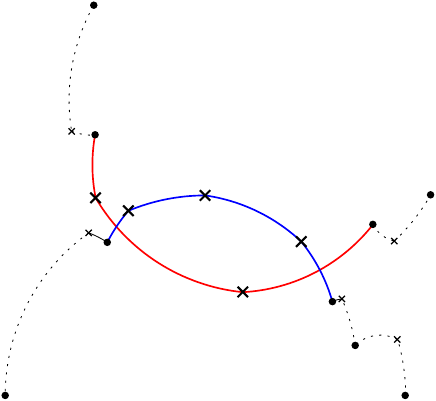}
  \caption{\label{fig:alternate} Ejemplo de cadenas intercaladas}
\end{figure}

Every intersection point will be associated with the respective
links. After sorting these intersection points lexicographically
according to the labels of the links involved in the intersection, we
will be able to track corresponding pairs of links (Figure
\ref{fig:alternate}), since this links correspond to a couple of
intersection points that involve the same links. We say that each of
this couples form an \emph{alternate link}.

For each alternate link let
$A_{\alpha}={\alpha_1, \alpha_2, \ldots, \alpha_v}$ be the set of red
arcs and $A_{\beta}={\beta_1, \beta_2, \ldots, \beta_w}$ the set of
blue arcs comprised in the alternate link.

Let us recall that an arc $\alpha_i$ is associated with an interval of
orientations $(\theta_{\alpha i_1},\theta_{\alpha i_2})$. We say that
$\theta_{\alpha i_1}$ is the \id{left} endpoint and
$\theta_{\alpha i_2}$ is the \id{right} endpoint of the
interval. Using these intervals we can verify if the conditions of
Lemma \ref{lem:overlap:conditions} are fulfilled. Bellow we describe
how the verification is done in $O(v+w)$ time.

Let us consider two functions: A boolean function \func{verifyOverlap}
that will accept a couple of arcs as parameters and will return
\const{true} if the conditions for Lemma \ref{lem:overlap:conditions}
are fullfilled by both arcs. And function \func{overlap} that takes
two arcs and returns the interval of orientations in which the overlap
takes place between the arcs. Each overlap interval, the corresponding
arcs and the points of $P$ that support the $\theta$-wedges that
overlap, are stores in an ordered list that we call \id{OverlapList}.

\begin{codebox}

\Procname{\proc{searchOverlap}(\id{A_{\alpha}}, \id{A_{\beta}})}
\li $i,m \gets 1$
\li \While $(i \leq \attribxx{A_{\alpha}}{\id{length}})$ \kw{ and } $(m \leq \attribxx{A_\beta}{length})$ 
\label{li:condicion-fin}
\li \Do 
	 \If  $(\attribxx{\alpha_i}{\id{right}} < \attribxx{\beta_m}{\id{left}})$
	 \label{li:intervalos-ajenos1}
\zi	\Then	 
	      \Comment disjoint intervals. Take next arc in \id{A_{\alpha}} 
\li	      $i \gets i+1$
\li	\ElseIf $(\attribxx{\beta_m}{\id{right}} < \attribxx{\alpha_i}{\id{left}})$
\label{li:intervalos-ajenos2}
 \zi	\Then	 
 	      \Comment disjoint intervals. Take next arc in \id{A_{\beta}} 
\li	      $m \gets m+1$
\li 	\ElseIf  $(\attribxx{\alpha_i}{\id{left}} < \attribxx{\beta_m}{\id{left}}$  \kw{ and }
               $\attribxx{\alpha_i}{\id{right}} > \attribxx{\beta_m}{\id{right}})$ 
	      \label{li:intervalos-caso2y6}
\zi	\Then	
        \Comment interval of \id{\beta_m} contained in \id{\alpha_i}'s interval 
\li		    \If $\func{verifyOverlap}(\id{\alpha_i},\id{\beta_m})$
\zi		    \Then
          \Comment Add overlap to list 
\li       \attribxx{\id{OverlapList}}{add}$(\id{\alpha_i}, \id{\beta_m}, \func{overlap}(\alpha_i, \beta_m) )$        
\zi       \End
          \Comment end if. Take next arc in $\id{A_{\beta}}$
\li    	$m \gets m+1$        

\li 	\ElseIf  $(\attribxx{\beta_m}{\id{left}} < \attribxx{\alpha_i}{\id{left}}$ \kw{ and } 
               $\attribxx{\beta_m}{\id{right}} > \attribxx{\alpha_i}{\id{right}})$ 
\zi   \Then
    	\Comment $\alpha_i$'s interval is contained in $\beta_m$'s. 
\li		    \If $\func{verifyOverlap}(\id{\alpha_i},\id{\beta_m})$
\zi		    \Then
          \Comment Add overlap to list 
\li       \attribxx{\id{OverlapList}}{add}$(\id{\alpha_i}, \id{\beta_m}, \func{overlap}(\alpha_i, \beta_m) )$        
\zi       \End
          \Comment end if. Take next arc in $\id{A_{\alpha}}$
\li    	$i \gets i+1$        
\li	  \ElseIf $(\attribxx{\beta_m}{\id{left}} < \attribxx{\alpha_i}{\id{left}})$ \kw{ and }
              $(\attribxx{\beta_m}{\id{right}} < \attribxx{\alpha_i}{\id{right}})$
	      \label{li:intervalos-caso3y5}
\zi	\Then	
     \Comment alternated intervals
\li		\If $\func{verifyOverlap}(\id{\alpha_i},\id{\beta_m})$
\zi   \Then
          \Comment Add overlap to list 
\li       \attribxx{\id{OverlapList}}{add}$(\id{\alpha_i}, \id{\beta_m}, \func{overlap}(\alpha_i, \beta_m) )$        
\zi       \End
          \Comment end if. Take next arc in $\id{A_{\beta}}$
\li		$m \gets m+1$
\li   \Else
\li		$i \gets i+1$
	\End 
\End 
\end{codebox}

\begin{lemma}
  Computational complexity of \proc{searchOverlap} is $O(v+w)$, where
  $v$ and $w$ are the number of arcs in the alternate link.
\end{lemma}

\begin{proof}
  The algorithm finishes when at least one of the chains has been
  completely traversed. On each iteration is its possible to leave
  behind one arc of one link. Functions \func{verifyOverlap} and
  \func{overlap} take constant time to execute each time.
\end{proof}

\section{Minimum area algorithm \label{sec:minimum-area}}

The list of event points obtained in the previous sections generate a
set of angular intervals $({\theta_1},{\theta_2})$, in which the set
of vertices of $\rcht{P}$ and the set of overlaps in $\rcht{P}$
remains unchanged.

\begin{figure}[ht!]
  \centering
  \subfloat[\label{fig:area:poly}The polygon]{\includegraphics[scale=0.5]{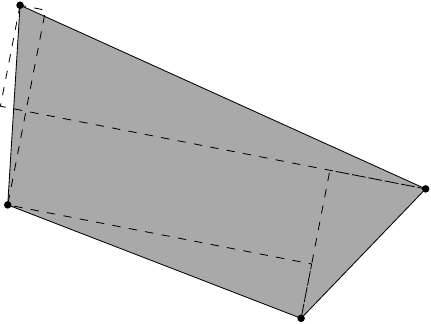}}
  \hspace{0.5cm}
  \subfloat[\label{fig:area:trian}]{\includegraphics[scale=0.5]{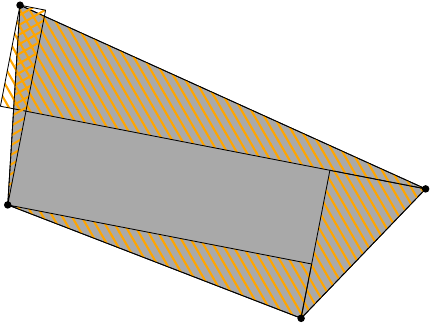}}
  \hspace{0.5cm}
  \subfloat[\label{fig:area:overlap}Out-event]{\includegraphics[scale=0.5]{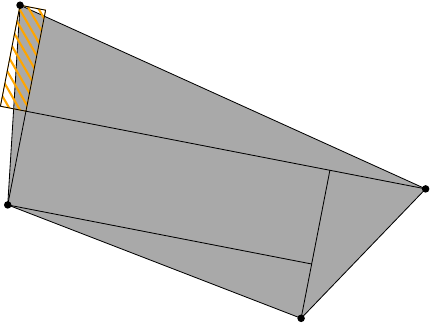}}
  \caption{Updating the area of the polygon \label{fig:area:polygon}}
\end{figure}

For any $\theta \in ({\theta_1},{\theta_2})$, the area of $\rcht{P}$
is given by the formula in~\cite{coreanos}. The formula has three main
components: the first is a polygon constructed by joining
counterclockwise consecutive vertices of $\rcht{P}$ (Figure
\ref{fig:area:poly}), the second is the sum of the areas of triangles
formed by consecutive vertices of $\rcht{P}$ and the apex of the
$P$-free $\theta$-wedge that they support (Figure
\ref{fig:area:trian}), finally the third one is the sum of the
overlaps area (Figure \ref{fig:area:overlap}).

\begin{figure}[ht!]
  \centering
  \subfloat[\label{fig:algor:area_in}In-event]{\includegraphics[scale=1]{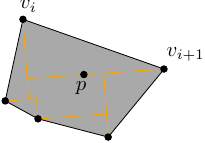}}
  \hspace{1cm}
  \subfloat[\label{fig:algor:area_middle}]{\includegraphics[scale=1]{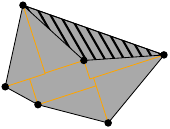}}
  \hspace{1cm}
  \subfloat[\label{fig:algor:area_out}Out-event]{\includegraphics[scale=1]{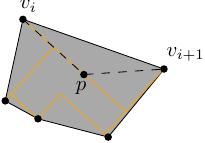}}
  \caption{Updating the area of the polygon \label{fig:algor:polygon}}
\end{figure}

It takes $O(n)$ time to compute the area of the polygon for the first
time. This polygon changes with each in- and out-event.  Let us say
that a point $p$ is getting in the $\rcht{P}$, between points
$v_i, v_{i+1}$ (see Figure \ref{fig:algor:area_in}). The area of the
polygon that includes $p$ can be computed in constant time from the
area of the previous polygon by subtracting the area of the triangle
$\triangle(v_i,p,v_{i+1})$ (Figure \ref{fig:algor:area_middle}). In
the out event of $p$, the area can be computed adding the area of the
same triangle (Figure \ref{fig:algor:area_out}).

Using the in and out-event list, it is possible to compute the set of
triangles that conform the second element of the formula. This list
has a linear number of changes, since on each in-event two triangles
are added to the list and a triangle is deleted form it and the same
amount of triangles changes on each out-event.

The areas of the overlaps can be computed in a way similar to the
triangles, but with the help of the ordered list of start- and
stop-events.

For each interval without changes, we update the corresponding values
of the formula in constant time by subtracting or adding new constant
values.

There can be more than one $\theta$-orientation in which $\rcht{P}$
has minimum area, but our algorithm is able to report them all.

From the discussion above, and the fact that the convex hull of $P$
can be computed from the rectilinear convex hull of $P$ in $O(n)$
time, we obtain the following:

\begin{theorem}
  Computing the set of orientations for which the rectilinear convex
  hull of $P$ has minimum area, can be done in optimal
  $\Theta(n\log n)$ time and $O(n)$ space.
\end{theorem}

\section{Concluding remarks}
\label{sec:concluding-remarks}

We have proved that the orientation in which $\rcht{P}$ has minimum
area can be obtained in optimal $O(n \log n)$ time, since with the
help of $O(n \log n)$ time preprocessing it is possible to maintain
$\rcht{P}$ as the \tor~changes from cero to $2\pi$.

Our result improves the quadratic algorithm given by Bae et
al.~\cite{coreanos}. It would be interesting to apply this results to
optimization problems in bichromatic sets, like finding the \tor~in
which the rectilinear convex hull of the red points is empty of blue
points or contains the minimum number of blue points, and the minimum
or maximum area rectilinear red convex hull, for example.

It is almost immediate to extend this algorithm to dynamic pointsets
and it would be very interesting to consider the problem of
maintaining the rectilinear hull of a moving pointset.

\end{document}